\documentclass[twocolumn,twocolappendix]{aastex701}
\shorttitle{Refractive Substructure in \txs}
\shortauthors{Plavin, Pushkarev, \& Kovalev}
\received{February 27, 2026}
\revised{April 7, 2026}
\accepted{April 16, 2026}
\submitjournal{ApJL}

\usepackage{xspace}
\usepackage{amsmath}
\usepackage{makecell}

\newcommand{\txs}{TXS\,2005+403\xspace}

\begin{document}

\title{Direct VLBI Detection of Interstellar Turbulence Imprint on a Quasar: \txs}

\correspondingauthor{A.~V.~Plavin}
\email{alexander@plav.in}

\author[0000-0003-2914-8554]{A.~V.~Plavin}
\email{alexander@plav.in}
\affiliation{Black Hole Initiative, Harvard University, 20 Garden St, Cambridge, MA 02138, USA}

\author[0000-0002-9702-2307]{A.~B.~Pushkarev}
\email{pushkarev.alexander@gmail.com}
\affiliation{Crimean Astrophysical Observatory, 298409 Nauchny, Crimea}

\author[0000-0001-9303-3263]{Y.~Y.~Kovalev}
\email{yykovalev@gmail.com}
\affiliation{Max-Planck-Institut f\"ur Radioastronomie, Auf dem H\"ugel 69, Bonn D-53121, Germany}

\begin{abstract}

We report the first unambiguous detection of refractive substructure in an active galactic nucleus (AGN) using ground-based Very Long Baseline Interferometry (VLBI). Our analysis of \txs~--- observed at 1--5~GHz along a line of sight through the Cygnus region --- reveals clear signatures of turbulence-induced substructure on long baselines that cannot be explained by the smooth scatter-broadened profile from diffractive effects alone. This signal persists across multiple observations spanning 2010--2019, demonstrating stable scattering properties along this line of sight. The combination of high flux density, compact intrinsic structure, and strong scattering establishes \txs as an exceptional laboratory for probing Galactic turbulence. This detection demonstrates that AGNs can serve as cosmic lighthouses illuminating interstellar plasma across the sky, complementing pulsar scintillation studies and informing scattering mitigation for millimeter-wavelength imaging of Sagittarius~A$^*$.

\end{abstract}


\section{Introduction} \label{s:intro}

Emission from extremely compact radio sources, such as active galactic nuclei (AGNs) with milliarcsecond (mas)-scale structure, is affected by scattering in the ionized interstellar medium of the Milky Way. Electron density fluctuations along the line of sight act as a turbulent ``screen'' that distorts the incoming radio waves. Diffractive scattering blurs compact sources into smooth, broadened profiles \citep{1976MNRAS.174....7D,1990ARA&A..28..561R}, while refractive scattering from turbulent density fluctuations imprints additional small-scale substructure \citep{1989MNRAS.238..963N,2015ApJ...805..180J,2016ApJ...826..170J,2016ApJ...833...74J}.

Radio Very Long Baseline Interferometry (VLBI), which achieves the highest angular resolution in astronomy by combining signals from telescopes separated by thousands of kilometers, is ideally suited to studying both effects. We previously measured diffractive broadening in large samples of AGNs and constructed all-sky scattering maps \citep{2015MNRAS.452.4274P,2022MNRAS.515.1736K}. Refractive substructure is far more challenging to detect. Substructure within scatter-broadened images was first observed in RadioAstron space VLBI studies of heavily scattered pulsars \citep{2016ApJ...822...96G,2017MNRAS.465..978P,2020ApJ...888...57P}. In the context of compact continuum sources, it was detected for Sagittarius~A$^*$ at 1.3~cm using the VLBA together with the Green Bank Telescope and interpreted as refraction by large-scale plasma fluctuations \citep{2014ApJ...794L..14G}, with subsequent observations and modeling \citep{2018ApJ...865..104J,2019ApJ...871...30I,2021ApJ...922L..28J}. For AGNs, RadioAstron has provided detections consistent with refractive substructure in 3C\,273 \citep{2016ApJ...820L..10J} and B\,0529+483 \citep{2018MNRAS.474.3523P}, but the interpretation is complicated by the degeneracy with intrinsic compact structure. AGNs provide background sightlines that sample the full Galactic column, unlike pulsars, which are nearby and concentrated toward the Galactic plane. Exploiting refractive scattering as a probe of the interstellar medium requires bright, heavily scattered AGNs, for which compact substructure is both detectable and not degenerate with the intrinsic emission.

One such object is the blazar \txs ($z = 1.736$; \citealt{1976MNRAS.177P..43B}), a bright (about 2~Jy) source behind the Cygnus region, where plasma clouds produce strong Galactic scattering (\autoref{a:skymap}; \citealt{2022MNRAS.515.1736K}). At frequencies $\gtrsim 15$~GHz, where scattering is subdominant, it displays a typical one-sided parsec-scale jet with superluminal apparent motion \citep{2023MNRAS.526.5932K,2021ApJ...923...30L}. At lower gigahertz frequencies, scattering along this line of sight was recently quantified through diffractive broadening measured with VLBI and through extreme scattering events in single-dish light curves \citep{2023MNRAS.526.5932K,2025MNRAS.542.2733K}; it was further discussed in \citet{2025ApJS..276...38P} as a limiting factor for astrometric accuracy.
The combination of high flux density and strong scattering makes \txs particularly well suited for probing interstellar plasma properties with VLBI.

Here, we report the first direct detection of refractive scattering substructure in an AGN using ground-based VLBI. We analyze archival Very Long Baseline Array (VLBA) observations of \txs at 1--5~GHz spanning nearly a decade (Section~\ref{s:data}), characterize anisotropic diffractive broadening (Section~\ref{s:diffractive}), and detect refractive substructure manifested as an otherwise unexplained long-baseline visibility signal (Section~\ref{s:detect}). The detected signal is consistent with interstellar turbulence models (Section~\ref{s:substructure_quant}). We summarize and discuss future prospects in Section~\ref{s:conclusions}.

\section{Data} \label{s:data}

\begin{table}[bt]
\begin{center}
\setlength{\tabcolsep}{4pt}
\footnotesize
\caption{VLBA long-baseline detections for \txs.}
\label{t:detsummary}
\begin{tabular}{lccccc}
\hline\hline
Exp. & Epoch & Freq. & BW & \multicolumn{2}{c}{Long-baseline detections} \\
code & & (GHz) & (MHz) & Per IF & Combined \\
(1) & (2) & (3) & (4) & (5) & (6) \\
\hline
BG196 & 2010-11-05 & 1.4 & 64 & 5\% & 10\% \\[0.7em]
BS224 & 2013-07-04 & 1.4 & 16 & 0\% & 4\% \\[0.7em]
BG246 & 2017-07-12 -- & 1.4 & 256 & 24\% & 54\% \\
 & -- 2018-01-08 & 2.3 & 256 & 14\% & 51\% \\
 & & 5.0 & 256 & 49\% & 86\% \\[0.7em]
UG002 & 2018-08-11 & 2.3 & 128 & 22\% & 59\% \\[0.7em]
BG258 & 2019-02-20 & 1.4 & 256 & 22\% & 56\% \\
 & & 2.3 & 256 & 14\% & 36\% \\
 & & 5.0 & 256 & 100\% & 100\% \\
\hline
\end{tabular}
\end{center}
\tablecomments{Columns are as follows: (1)~--~experiment code; (2)~--~observation epochs for the experiment as YYYY-MM-DD; (3)~--~nominal frequency; (4), BW~--~total bandwidth (per polarization); (5)~--~percentage of long-baseline detections from individual IF fringe searches; (6)~--~percentage of detections when adjacent IFs are combined. Long-baseline detections are defined as signal detected on baselines beyond where the diffractive-broadening Gaussian model reaches 1/1000 of its peak amplitude. See \autoref{a:fft} for details on signal search procedure.}
\end{table}

Our analysis focuses on 1--5~GHz, where refractive and diffractive scattering effects from the interstellar medium are most pronounced.
We use Very Long Baseline Array (VLBA) observations from 2010--2019 available in the NRAO data archive, experiment codes BG196, BG246, BG258, BS224, and UG002. A summary of these observations is provided in \autoref{t:detsummary}.
The observations used 16--256~MHz total bandwidth (per feed polarization) split into 4--8 intermediate-frequency subbands (hereafter IFs), depending on the experiment. All datasets were calibrated using standard \textit{AIPS} \citep{2003ASSL..285..109G} procedures for VLBI data reduction.

\section{Results} \label{s:results}

\begin{figure*}[ht]
\includegraphics[width=0.32\linewidth]{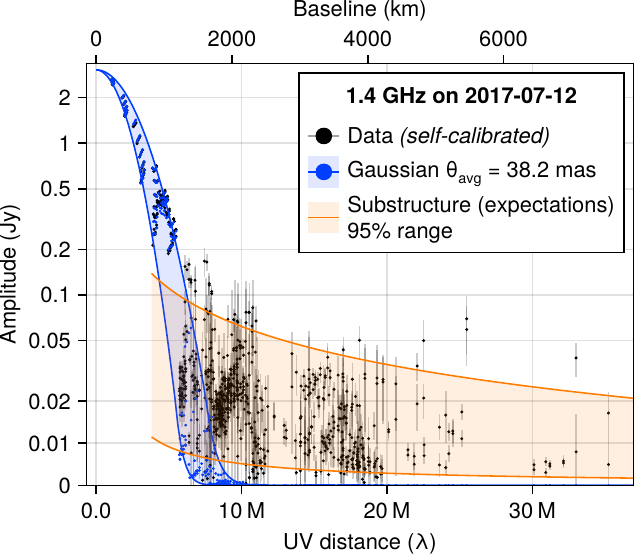}
\includegraphics[width=0.32\linewidth]{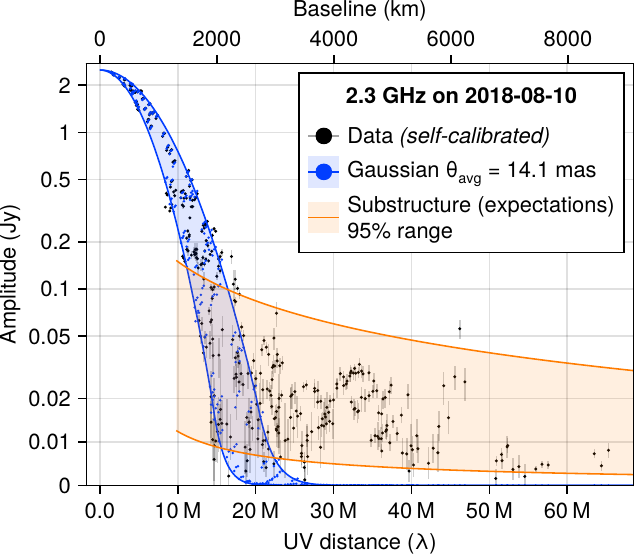}
\includegraphics[width=0.32\linewidth]{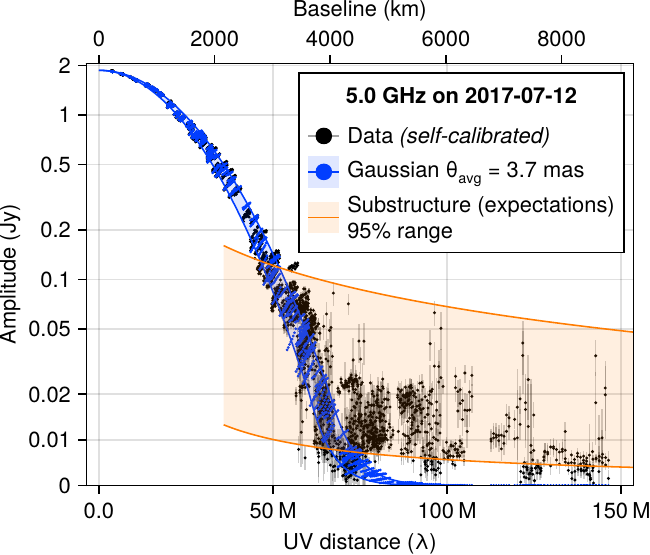}
\caption{Visibility amplitude versus projected baseline length or $uv$-distance. Experiments with the best sensitivity and coverage are shown in each frequency band, left to right: 1.4, 2.3, 5~GHz. Data (black dots with $1\sigma$ error bars) are self-calibrated to per-IF elliptical Gaussian models (blue dots). The blue shaded band shows the average Gaussian model within each band. The predicted refractive substructure signal is shown as the orange band starting from the $uv$-distance where the Gaussian model first falls to 25\% of its peak. Its calculations assume scattering with broadening size equal to the average axis of the Gaussian (see \autoref{s:substructure_quant}).}
\label{f:radplots}
\end{figure*}

\subsection{Anisotropic Diffractive Broadening} \label{s:diffractive}

\begin{figure}[ht]
\centering
\includegraphics[width=\linewidth]{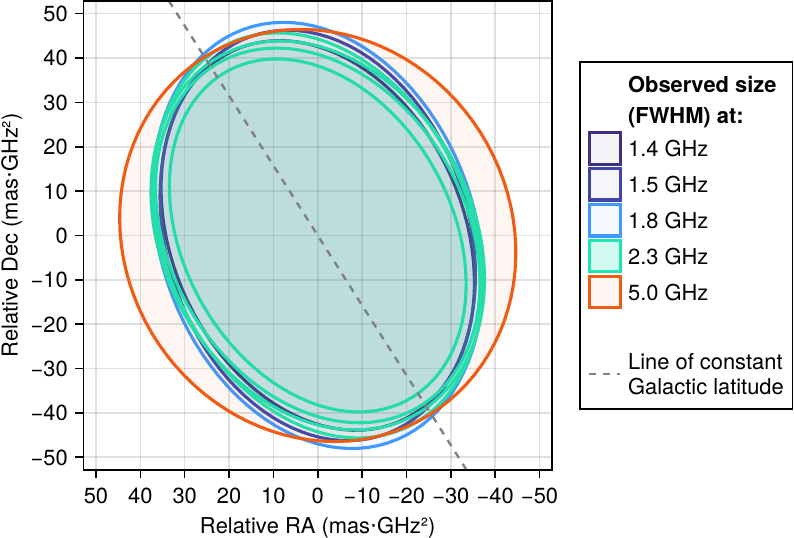}
\vspace{0.01em}\\
\includegraphics[width=\linewidth]{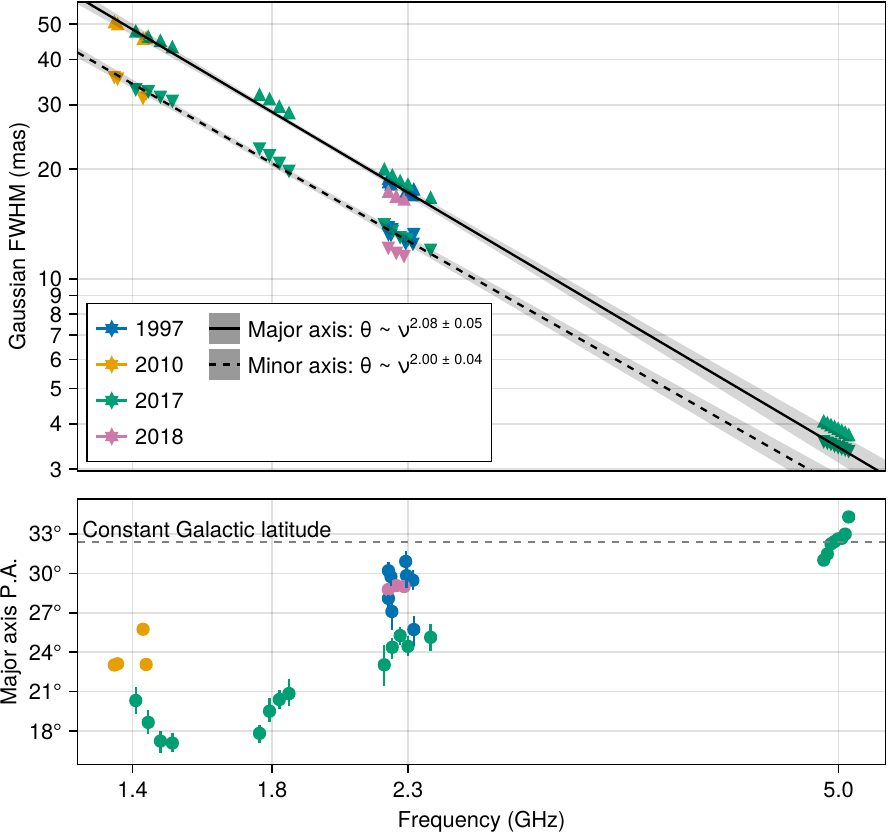}
\caption{Elliptical Gaussian fits to VLBA observations at 1--5~GHz.\\ \textit{Top}: Gaussian contours at the half-maximum level (FWHM) in RA--Dec space; each ellipse represents a single observation, with sizes normalized by $\nu^2$ for direct comparison across frequencies.\\ \textit{Middle and bottom}: Frequency dependence of Gaussian major and minor axes (middle) and position angle (bottom); each point represents a single IF with $2\sigma$ uncertainties shown.\\ All panels show clear $\approx\nu^{-2}$ size scaling (5~GHz is excluded from the powerlaw fit) and consistent elongation along the Galactic plane.}
\label{f:broadening_size_vis}
\end{figure}

As demonstrated in \cite{2023MNRAS.526.5932K}, the morphology of \txs is dominated by its parsec-scale jet at $\gtrsim 8$~GHz and by scatter broadening at $\lesssim 5$~GHz. Here, we focus on the latter frequency range (1--5~GHz).

To characterize diffractive broadening, we fit a single elliptical Gaussian component to each frequency band in every experiment. We use only short baselines with a signal-to-noise ratio (SNR) above 10 and fit closure amplitudes to avoid relying on a priori calibration. The visibilities are then self-calibrated (both amplitude and phase of the gains) to these per-IF Gaussian models for plotting and further analysis.

The elliptical Gaussian model shows excellent agreement with the short-baseline data at 1--5~GHz (\autoref{f:radplots}). The fitted sizes (\autoref{f:broadening_size_vis}) follow $1/\nu^2$ scaling, much steeper than intrinsic core-size frequency dependence and consistent with the scattering-dominated regime \citep{1990ARA&A..28..561R}. At 5~GHz, intrinsic source structure likely contributes, causing a slight size increase~--- particularly along the minor axis, where scattering is weaker. The minor axis also lies close to the inner jet direction \citep{2022ApJS..260....4P,2023MNRAS.526.5932K}, so we exclude the 5~GHz band from the power-law fit. The fitted ellipse shape remains consistent across epochs and frequencies, with elongation aligned along the Galactic plane. This alignment was first reported for another blazar (B2\,2023+335) in \citet{2013A&A...555A..80P}, then detected for \txs in \citet{2023MNRAS.526.5932K}, and was interpreted to indicate persistent structures in the plasma screen aligned perpendicular to the Galactic plane. In what follows, we use these Gaussian models of the diffractive broadening to distinguish refractive substructure from potential intrinsic source structure contribution.

\subsection{Refractive Substructure Detection} \label{s:detect}

\begin{figure*}[!ht]
    \centering
    \includegraphics[height=0.32\linewidth]{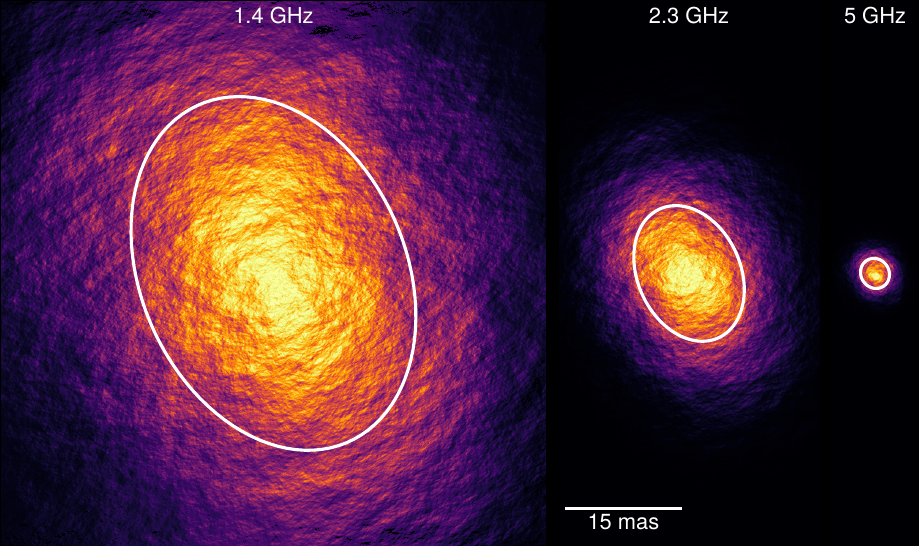}
    \includegraphics[height=0.32\linewidth]{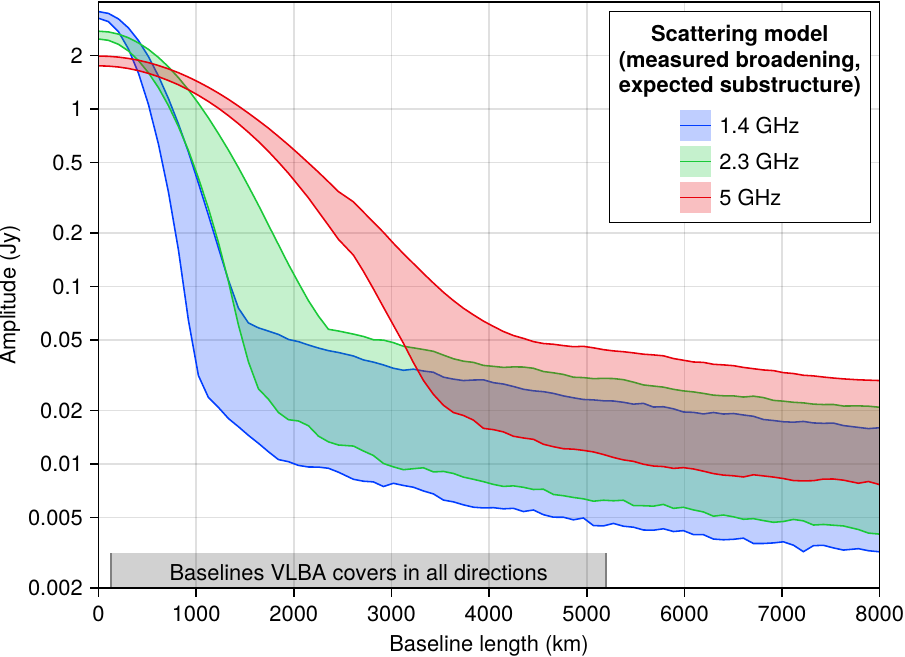}
    \caption{Simulated scattering of \txs using parameters consistent with our observations (\autoref{s:substructure_quant}).\\
    \textit{Left}: a single realization of the scattered image at 1.8, 2.3, and 5~GHz, showing both the large-scale Gaussian broadening and fine-scale refractive substructure from turbulent density fluctuations in the scattering screen. White ellipses mark the measured FWHM sizes from elliptical Gaussian fits (\autoref{f:broadening_size_vis}), averaged per band.\\
    \textit{Right}: visibility amplitude predictions; shaded bands show the 95\% range across random screen realizations. The highlighted baseline range indicates where the VLBA provides good position angle coverage.\\
    An interactive version of this figure is available at \url{https://github.com/aplavin/txs2005-refractive-substructure}.}
    \label{f:simimage}
\end{figure*}

The Gaussian models describe the observed large-scale structure well. However, as shown in \autoref{f:radplots}, we observe significant signal on long baselines well beyond where the Gaussian envelope predicts any measurable flux. The long-baseline signal exhibits fluctuations correlated in the $uv$ plane, supporting the presence of coherent structures and not observational noise. This excess cannot arise from intrinsic blazar structure: diffractive scattering acts as a convolution, suppressing the observed visibility by a baseline-dependent factor --- the Fourier transform of the scattering kernel, well described by a Gaussian (\autoref{s:diffractive}). At baselines where this factor drops below $\sim10^{-3}$ of its peak, any intrinsic emission is attenuated to negligible levels regardless of its morphology. This leaves refractive substructure from turbulence in the interstellar medium as the only mechanism capable of producing signal beyond the broadening envelope (see \autoref{f:simimage} for an illustration). The strong scattering toward \txs creates a clean separation in the $uv$ plane: diffractive broadening dominates on short baselines, while only refractive substructure contributes on long baselines --- a regime distinction not accessible for the AGNs studied earlier.

We detect the signal shown in \autoref{f:radplots} using standard \textit{AIPS} antenna-based fringe fitting and independently confirm it with a baseline-based FFT search applied directly to the raw correlator data, bypassing any preprocessing and avoiding potential SNR biases. Both approaches yield consistent results with numerous robust detections on long baselines. For motivation and full details of the fringe search methodology and results, see \autoref{a:fft}.

A summary of long-baseline signal detections is given in \autoref{t:detsummary} and shown in \autoref{f:detsummary}. A long-baseline signal incompatible with pure diffractive broadening is confidently detected across all three frequency bands (1, 2, 5~GHz) from 2010 to 2019. We therefore conclude that refractive substructure in \txs has been detected using ground-based VLBI --- for the first time in any AGN. The substructure effects persist over nearly a decade (\autoref{f:dettime}).

\subsection{Comparison with Turbulence Models} \label{s:substructure_quant}

Refractive substructure probes density fluctuations at scales comparable to the scattering disk projected onto the screen \citep{1989MNRAS.238..963N,2015ApJ...805..180J}. For the screens in the Cygnus region at $D \approx 1.4$~kpc \citep{2012A&A...539A..79R}, this corresponds to a few AU to tens of AU (1~mas $\approx$ 1.4~AU). Detecting refractive scattering shows that the interstellar medium has structure at these large scales --- a signature of turbulence with a broad spatial power spectrum, such as a power law.

The substructure introduces a fundamentally stochastic signal in the visibilities. We compare the \cite{2015ApJ...805..180J} model to the observed long-baseline amplitudes (\autoref{s:data}), computing expected visibility amplitude variations from a power-law turbulence scale spectrum with $\alpha = 5/3$ and an inner scale $r_{\text{in}} = 1000$~km. The data generally fall within the predicted range (\autoref{f:radplots}); \autoref{f:simimage} shows simulated images with substructure based on these parameters. Given the limitations of the available data, especially the effective noise level, we perform only this consistency check and do not attempt to constrain the turbulent screen properties.

The turbulence inner scale can also be probed through the diffractive broadening profile, which is expected to depart from a Gaussian at baselines exceeding the inner scale \citep[see, e.g.][]{2018ApJ...865..104J}. We find no clear evidence for this effect in our data, suggesting that $r_\mathrm{in}$ lies beyond the baselines where the scattering kernel visibility drops well below its peak, i.e.\ $\gtrsim 1000$~km.

With its strong scattering and high flux density, \txs is a prime candidate for probing these turbulence properties using both broadening and substructure observables. The two scattering regimes are well-separated by baseline length for this line of sight at gigahertz frequencies, and are readily accessible to ground-based VLBI arrays. We plan to pursue this with our ongoing VLBA campaign, together with a more detailed substructure variability analysis.

The turbulent scattering models indicate that the 1--10~GHz VLBI regime is optimal for heavily scattered AGNs like \txs: broadening dominates the apparent structure while refractive scattering produces measurable long-baseline correlated signal. The refractive visibility fluctuations are expected to be even larger at higher frequencies (see the right panel of \autoref{f:simimage}) on baselines accessible to space VLBI, but distinguishing that signal from intrinsic compact structure becomes harder, as was seen in RadioAstron AGN results \citep{2016ApJ...820L..10J,2018MNRAS.474.3523P}.

\section{Summary} \label{s:conclusions}

\txs is a bright ($\sim$2~Jy) blazar viewed through the Cygnus region, where Galactic scattering is particularly strong. We detect a long-baseline signal in its VLBI observations that cannot be explained by the diffractive scatter-broadened profile or by intrinsic source structure. We identify this signal as refractive substructure produced by turbulence in the intervening plasma, making \txs the first AGN in which refractive scattering has been directly detected with ground-based VLBI. This signal is detected from 1.4 to 5~GHz and remains stable across nearly a decade, with amplitudes consistent with theoretical expectations. This consistency indicates a persistently strong scattering screen with stable properties along this line of sight.

\begin{figure*}[t]
\centering
\includegraphics[width=0.8\linewidth]{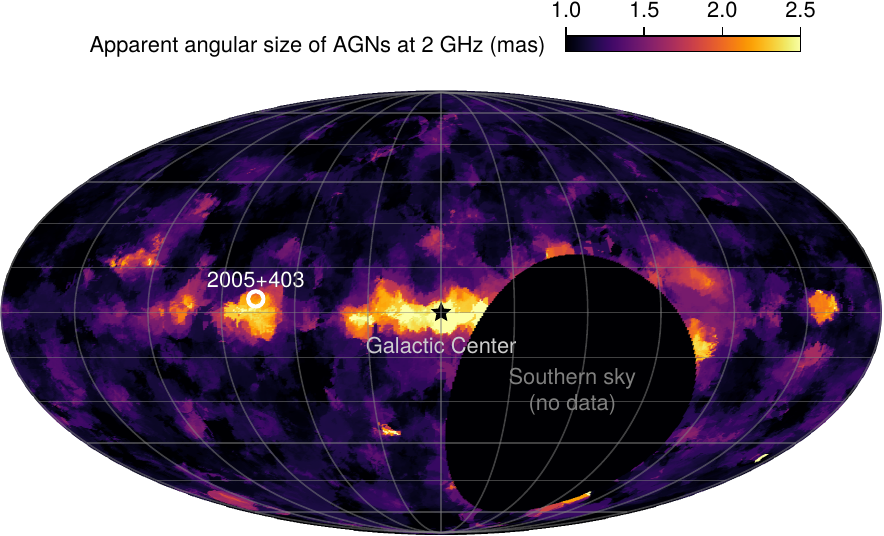}
\caption{All-sky map of the apparent size of 3698 AGNs measured with VLBI at 2~GHz, as a proxy for interstellar scattering strength, following \citet{2022MNRAS.515.1736K}. The position of \txs is highlighted, located behind the Cygnus region --- one of the strongest scattering environments in the Galaxy.}
\label{f:skymap}
\end{figure*}

These findings establish \txs as an exceptional laboratory for studying Galactic turbulence and scattering physics. The source combines several crucial properties: (i)~high flux density ($\sim$2~Jy), enabling detections with routine VLBI; (ii)~compact intrinsic structure on mas and sub-mas scales, necessary for scattering to dominate the observed morphology; (iii)~structural stability on timescales of months, unlike Sagittarius~A$^*$, where intrinsic variability complicates interpretation; and (iv)~strong scattering due to its location behind the turbulent Cygnus region. This detection suggests that similar AGNs in other strongly scattering regions could be identified (see \autoref{f:skymap}), enabling systematic studies of Galactic turbulence and magnetic field structure across the sky. Improved understanding of scattering properties from sources like \txs would directly inform efforts to mitigate scattering artifacts in Event Horizon Telescope images of the black hole in the center of the Milky Way, where scattering limits image fidelity, and would help interpret propagation effects in fast radio bursts.

Dedicated VLBI observations with higher sensitivity and more uniform $uv$ coverage can move this result from detection to parameter inference. They will enable direct tests of turbulence models by jointly constraining the inner scale, anisotropy, and magnetic field-aligned structure in the scattering plasma. Multi-epoch time-domain measurements can quantify refractive modulation amplitudes and characteristic variability timescales, constraining the kinematics and evolution of the Cygnus scattering screen along this line of sight. We have initiated VLBA project BP276 with multi-epoch, multi-frequency observations in 2025--2026 to pursue these goals. Together with the future identification of analogous sources behind strongly scattering Galactic regions, such observations are establishing AGNs as a systematic probe of interstellar turbulence.

\begin{acknowledgments}

We thank Michael Johnson, Ramesh Narayan, Aditya Tamar, and an anonymous referee for useful comments on the manuscript.
AVP is a postdoctoral fellow at the Black Hole Initiative, which is funded by grants from the John Templeton Foundation (grants 60477, 61479, 62286) and the Gordon and Betty Moore Foundation (grant GBMF-8273). 
ABP was supported by the Russian Science Foundation grant 25-22-00152.
YYK was supported by the MuSES project, which has received funding from the European Union (ERC grant agreement No 101142396).
%
The views and opinions expressed in this work are those of the authors and do not necessarily reflect the views of these Foundations.
This work made use of results of the Swinburne University of Technology software correlator DiFX \citep{DiFX}, developed as part of the Australian Major National Research Facilities Programme and operated under license.

Code to reproduce analysis and figures from this manuscript can be found at \dataset[https://github.com/aplavin/txs2005-refractive-substructure]{https://doi.org/10.5281/zenodo.19464371}.

\end{acknowledgments}





\facilities{VLBA}

\software{AIPS \citep{2003ASSL..285..109G},
          Makie.jl \citep{Danisch2021},
          Pigeons.jl \citep{Surjanovic2024},
          ScatteringOptics.jl (\url{https://github.com/EHTJulia/ScatteringOptics.jl}),
          ScatteringOpticsExtra.jl (\url{https://github.com/JuliaAPlavin/ScatteringOpticsExtra.jl}),
          VLBIData.jl (\url{https://github.com/JuliaAPlavin/VLBIData.jl}),
          AccessibleModels.jl (\url{https://github.com/JuliaAPlavin/AccessibleModels.jl});
          a complete list of dependencies is available in the accompanying code repository.
          }

\appendix
\section{TXS\,2005+403 in the Context of All-Sky Scattering} \label{a:skymap}

The apparent angular sizes of AGNs observed with VLBI provide a direct measure of interstellar scattering strength across the sky. \autoref{f:skymap} shows such a map for 3698 sources at 2~GHz, the lowest frequency at which enough VLBI data are available. We follow the methodology of \citet{2022MNRAS.515.1736K}, incorporate data accumulated since that work, and apply more stringent filtering. The Cygnus region stands out as one of the strongest scattering environments accessible along AGN sightlines. \txs lies directly behind it, combining high flux density with extreme scattering --- the properties that make the refractive substructure detection reported in this work possible. This map also illustrates the potential for identifying analogous sources behind other strongly scattering Galactic regions.

\section{Fringe Search Methodology} \label{a:fft}

\begin{figure}[ht]
\centering\includegraphics[width=1\linewidth]{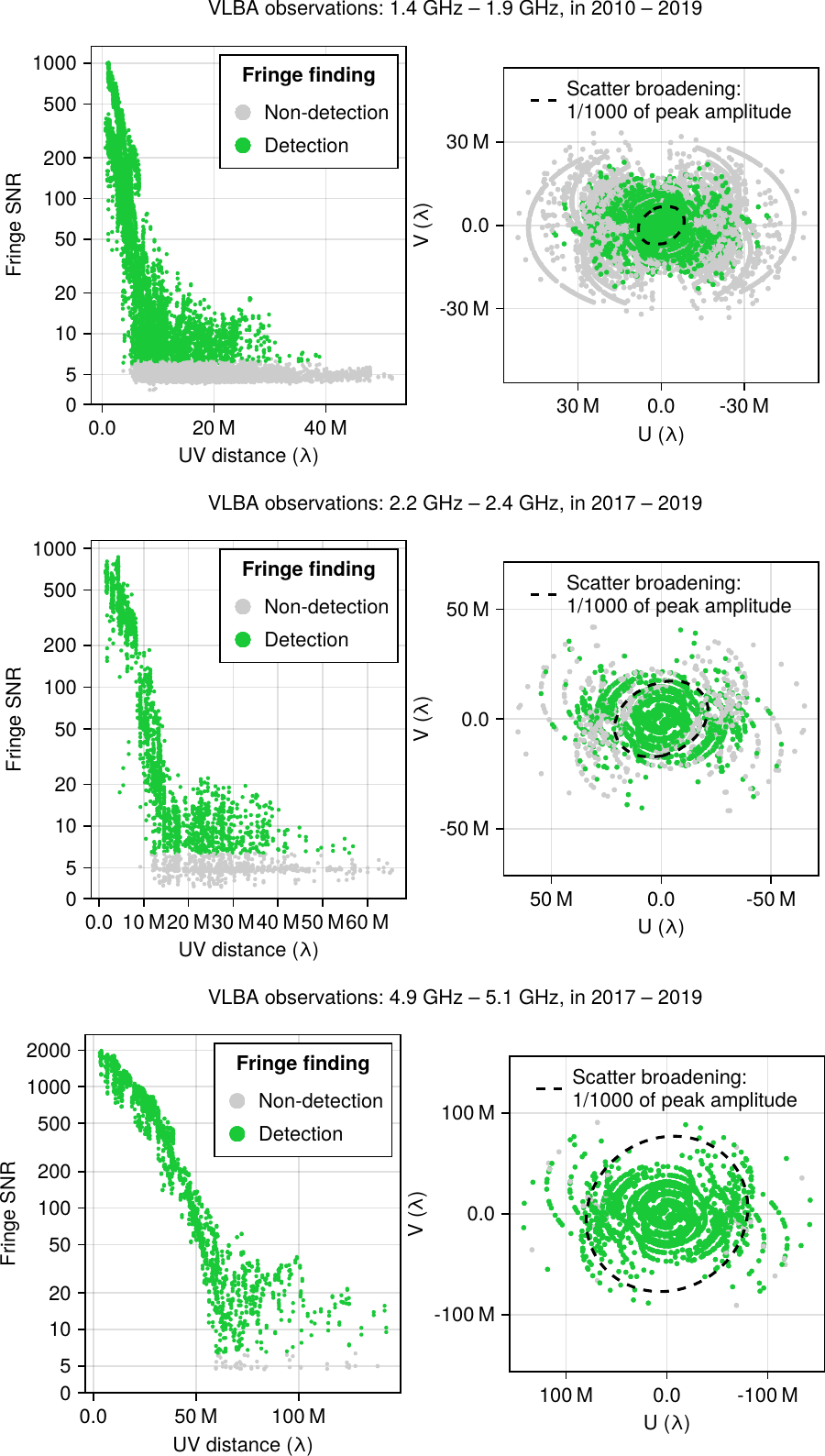}
\caption{Long-baseline detections in VLBA observations of \txs over 2010--2019. Individual panels show detections at 1, 2, and 5~GHz when adjacent IFs are combined, displaying their SNR and $uv$-plane locations. Elliptical contours in the $uv$ plane mark where the diffractive-broadening Gaussian reaches $1/1000$ of its peak --- a conservative boundary beyond which no signal is expected from pure scatter broadening. See \autoref{a:fft} and \autoref{t:detsummary} for detection details and threshold definitions.}
\label{f:detsummary}
\end{figure}

\begin{figure}[ht]
\vspace{1em}
\centering\includegraphics[width=0.8\linewidth]{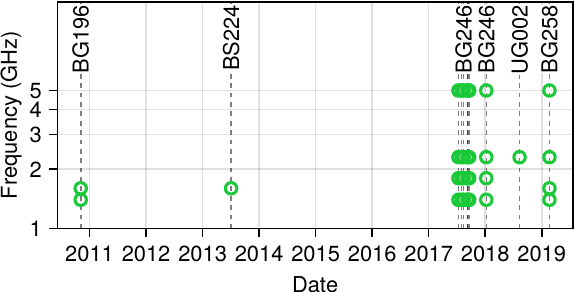}
\caption{Times and frequencies of all VLBA observations of \txs with detected long-baseline signal, spanning about a decade.}
\label{f:dettime}
\end{figure}

The visibility data shown in \autoref{f:radplots} are obtained from \textit{AIPS} processing with antenna-based fringe fitting performed independently for each IF. We require a baseline SNR cutoff of 5 and a minimum of three antennas for a detection. \textit{AIPS} yields numerous detections on baselines longer than expected from the Gaussian structure.

These visibilities exceed the commonly used SNR threshold, but only by a factor of a few. To be as conservative as possible at this detection edge, we implement an independent baseline-based fringe search directly on the raw correlator data from the VLBA archive, bypassing any preprocessing. This further avoids relying on \textit{AIPS} SNR estimates, which can be biased near the detection threshold \citep[see discussion in][]{2011AJ....142...35P,2020AdSpR..65..705K,2023A&A...676A.114S}. We compute the Fast Fourier Transform (FFT) of each scan and IF without restricting the search window. This search yields a $\pm1$~$\mu$s delay range (defined by the spectral-channel spacing within each IF) and a $\pm0.1$--$0.3$~Hz rate range (depending on the correlator integration time). As expected, individual FFT amplitudes follow Gaussian statistics in scans without any detected signal. All FFT output values are independent for a pure noise input without oversampling, and the spurious detection probability is $\mathbb{P}(\text{peak}|\text{noise}) \leq N e^{-\text{SNR}^2 / 2}$, where $N$ is the number of search grid cells \citep{2011AJ....142...35P}. We use $2\times$ oversampling in delay and rate to reduce peak amplitude loss from $\times2.5$ to $\times1.2$ \citep{2011AJ....142...35P}. Oversampling introduces dependencies between FFT values, so the output contains fewer than $N$ independent values. We conservatively retain the same probability formula, which becomes an upper bound. We adopt $\mathbb{P}(\text{peak}|\text{noise}) < 5\times10^{-5}$ as the conservative detection threshold, corresponding to fewer than 0.5 expected spurious detections across all observations. We first search each IF individually, then concatenate the raw data from IFs adjacent in frequency and repeat the fringe search over their total bandwidth to increase sensitivity. The resulting detections are summarized in \autoref{t:detsummary}, \autoref{f:detsummary}, and \autoref{f:dettime}.

\bibliography{scat2005403}
\bibliographystyle{aasjournalv7}

\end{document}